\documentclass{aa}
\usepackage[varg]{txfonts}
\usepackage{natbib, aas_macros}
\usepackage{graphicx} 
\usepackage{amsmath}

\begin{document}

\title{The Effect of Multiple Heat Sources on Exomoon Habitable Zones}

\author{Vera Dobos\inst{1,2}
 \and Ren\'e Heller\inst{3}
 \and Edwin L. Turner\inst{4,5}}


\offprints{V. Dobos, \email{dobos@konkoly.hu}}

\institute{Konkoly Thege Mikl\'os Astronomical Institute, Research Centre for Astronomy and Earth Sciences, Hungarian Academy of Sciences, H--1121 Konkoly Thege Mikl\'os \'ut 15-17, Budapest, Hungary
  \and Geodetic and Geophysical Institute, Research Centre for Astronomy and Earth Sciences, Hungarian Academy of Sciences, H--9400 Csatkai Endre u. 6-8., Sopron, Hungary
  \and Max Planck Institute for Solar System Research, Justus-von-Liebig-Weg 3, 37077 G\"ottingen, Germany
  \and Department of Astrophysical Sciences, Princeton University, 08544, 4 Ivy Lane, Peyton Hall, Princeton, NJ, USA
  \and The Kavli Institute for the Physics and Mathematics of the Universe (WPI), University of Tokyo, 227-8583, 5-1-5 Kashiwanoha, Kashiwa, Japan}

\date{Received 1 June 2015 / Accepted 1 June 2015}

\abstract
{With dozens of Jovian and super-Jovian exoplanets known to orbit their host stars in or near the stellar habitable zones, it has recently been suggested that moons the size of Mars could offer abundant surface habitats beyond the solar system. Several searches for such exomoons are now underway, and the exquisite astronomical data quality of upcoming space missions and ground-based extremely large telescopes could make the detection and characterization of exomoons possible in the near future. Here we explore the effects of tidal heating on the potential of Mars- to Earth-sized satellites to host liquid surface water, and we compare the tidal heating rates predicted by tidal equilibrium model and a viscoelastic model. In addition to tidal heating, we consider stellar radiation, planetary illumination and thermal heat from the planet. However, the effects of a possible moon atmosphere are neglected. We map the circumplanetary habitable zone for different stellar distances in specific star-planet-satellite configurations, and determine those regions where tidal heating dominates over stellar radiation. We find that the `thermostat effect' of the viscoelastic model is significant not just at large distances from the star, but also in the stellar habitable zone, where stellar radiation is prevalent. We also find that tidal heating of Mars-sized moons with eccentricities between 0.001 and 0.01 is the dominant energy source beyond 3--5 AU from a Sun-like star and beyond 0.4--0.6 AU from an M3 dwarf star. The latter would be easier to detect (if they exist), but their orbital stability might be under jeopardy due to the gravitational perturbations from the star.}

\keywords{astrobiology -- methods: numerical -- planets and satellites: general -- planets and satellites: interiors}
\maketitle

\section{Introduction}

Although no exomoon has been discovered as of today, it has been shown that Mars-mass exomoons could exist around super-Jovian planets at Sun-like stars, from a formation point of view \citep[][]{helleretal14,heller15,2015ApJ...806..181H}. If they exist, these moons could be detectable with current or near-future technology \citep{kippingetal09,heller14a,2015ApJ...810...29H}\,and they would be potentially habitable \citep{1997Natur.385..234W,heller14a,lammer14}.

For both planets and moons, it is neither sufficient nor even necessary to orbit their star in the habitable zone (HZ) -- the circumstellar region where the climate on an Earth-like planet would allow the presence of liquid surface water \citep{kasting93}. Instead, tidal heating can result in a suitable surface temperature for liquid water at large stellar distances \citep{reynolds87,peters13, 2014AsBio..14...50H, dobos15}. Moreover, most of the water -- and even most of the liquid water -- in the solar system can be found beyond the location of the past snowline around the young Sun \citep{kereszturi10}, which is supposed to have been at about 2.7\,AU during the final stages of the solar nebula \citep{1981PThPS..70...35H}. Hence, these large water reservoirs could be available for life on the surfaces of large exomoons at wide stellar separations.

However, observations of moons orbiting at several AU from their stars cannot be achieved with the conventional transit method, because transit surveys can only detect exomoons with significant statistical certainty after multiple transits in front of their star. This implies relatively small orbital separations from the star, typically $<1$\,AU \citep{2006A&A...450..395S}. It might, however, be possible to image exomoons with extreme tidal heating \citep{peters13} or to observe the transits of moons around luminous giant planets \citep{2014ApJ...796L...1H,2016A&A...588A..34H,2016ApJ...824...76S}\,at several AU from their host stars.

At this wide a distance from its host star, tidal heating in a large exomoon would be key to longterm surface habitability. Tidal heating of exomoons is usually calculated from equilibrium tide models, e.g. through parameterization with a tidal dissipation factor ($Q$) and with a uniform rigidity of the rocky material of the moon ($\mu$), both of which are considered constants. This family of the tidal equilibrium models is called the  ``constant-phase-lag'' (CPL) models because they assume a constant lag of the tidal phases between the tidal bulge of the moon and the line connecting the moon and its planetary perturber \citep{2013ApJ...764...26E}. Alternatively, equilibrium tides can be described by a ''constant-time-lag`` (CTL) model, where it is assumed that the tidal bulge of the moon lags the line between the two centers of mass by a constant time.

In reality, however, all these parameters depend on temperature. Another problem with equilibrium tide models is that the values of the parameters (e.g. of $Q$) are difficult to calculate \citep{2012A&A...541A.165R} or constrain observationally even for solar system bodies: $Q$ can vary several orders of magnitude for different bodies: from $\approx~10$ for rocky planets to $>~1000$ for giant planets \citep[see e.g.][]{goldreich66}.

For these reasons, \citet{dobos15} applied a viscoelastic model previously developed by \citet{henning09} to predict the tidal heating in hypothetical exomoons. Viscoelastic models are more realistic than tidal equilibrium models, as they consider the temperature feedback between the tidal heating and the viscosity of the material. Due to the phase transition at the melting of rocky material, viscoelastic models result in more moderate temperatures than do tidal equilibrium models, and so they predict a wider circumplanetary habitable zone \citep{forgan16}.

In this work, we compare the effects of two kinds of tidal heating models, a CTL model and a viscoelastic model, on the habitable edge for moons within the circumstellar HZ. We also calculate the contribution of viscoelastic model to the total energy flux of hypothetical exomoons with an emphasis on moons far beyond the stellar habitable zone.

\section{Method} \label{method}

\subsection{Energy flux at the moon's top of the atmosphere}

We neglect any greenhouse or cloud feedbacks by a possible moon atmosphere and rather focus on the global energy budget. The following energy sources are included in our model: stellar irradiation, planetary reflectance, thermal radiation of the planet, and tidal heating. The globally averaged energy flux on the moon is estimated as per \citep[Eq. 4]{helleretal14}

\begin{equation}
   \label{Fglob}
      \overline{F}_\mathrm{s}^\mathrm{glob} = \frac { L_* \left( 1 - \alpha_\mathrm{s,opt} \right) } { 4 \pi a_\mathrm{p}^2 \sqrt{1 - e_\mathrm{p}^2} } \left( \frac { x_\mathrm{s} } { 4 } + \frac {\pi R_\mathrm{p}^2 \alpha_\mathrm{p}} { f_\mathrm{s} 2 a_\mathrm{s}^2 } \right)+ \frac {L_\mathrm{p} \left( 1 - \alpha_\mathrm{s,IR} \right) } { 4 \pi a_\mathrm{s}^2 f_\mathrm{s} \sqrt{1 - e_\mathrm{s}^2} } + h_\mathrm{s} + W_\mathrm{s} \, ,
\end{equation}

\noindent where $L_*$ and $L_\mathrm{p}$ are the stellar and planetary luminosities, respectively, $\alpha_\mathrm{p}$ is the Bond albedo of the planet, $\alpha_\mathrm{s,opt}$ and $\alpha_\mathrm{s,IR}$ are the optical and infrared albedos of the moon, $a_\mathrm{p}$ and $a_\mathrm{s}$ are the semi-major axes of the planet's orbit around the star and of the moon's orbit around the planet\footnote{We assume that the moon's mass is negligible compared to the planetary mass and that the barycenter of the planet-moon system is in the center of the planet.}, $e_\mathrm{p}$ and $e_\mathrm{s}$ are the eccentricities of the planet's and the satellite's orbit, $R_\mathrm{p}$ is the radius of the planet, $x_\mathrm{s}$ is the fraction of the satellite's orbit that is not spent in the shadow of the host planet, and $f_\mathrm{s}$ describes the efficiency of the flux distribution on the surface of the satellite. The first and second terms of this equation account for illumination effects from the star and the planet, whereas $h_\mathrm{s}$ indicates the tidal heat flux through the satellite's surface. $W_\mathrm{s}$  denotes arbitrary additional energy sources such as residual internal heat from the moon's accretion or heat from radiogenic decays, but we use $W_\mathrm{s}=0$. We also neglect eclipses, thus $x_\mathrm{s}=1$, and we assume that the satellite is not tidally locked to its host planet, hence $f_\mathrm{s}=4$. The planetary radius is calculated from the mass, using a polynomial fit to the data given by \citet[Table 4, line 17]{fortney07}. The albedos of both the planet and the moon (in the optical and also in the infrared) were set to 0.3, that is, to Earth-like values.

We estimate the runaway-greenhouse limit of the globally averaged energy flux ($F_\mathrm{RG}$), which defines the circumplanetary habitable edge interior to which a moon becomes uninhabitable, using a semi-analytic model of \citet{2010ppc..book.....P} as described in \citet{heller13}. In this model, the outgoing radiation on top of a water-rich atmosphere is calculated using an approximation for the wavelength-dependent absorption spectrum of water. The runaway greenhouse limit then depends exclusively on the moon's surface gravity (i.e. its mass and radius) and on the fact that there is enough water to saturate the atmosphere with steam. Our test moons are assumed to be rocky bodies with masses between the mass of Mars (0.1 Earth masses, $M_\oplus$) and $1\,M_\oplus$. Our test host planets are gas giants around either a sun-like star or an M dwarf star.

We apply the model of \citet{kopparapu14} to calculate the borders of the circumstellar HZ, which are different for the 0.1 and the 1 Earth-mass moons. The corresponding energy fluxes are then used to define and evaluate the habitability of our test moons, both inside and beyond the stellar HZ.

\subsection{Tidal heating models}

Tidal heat flux is calculated using two different models: a viscoelastic one with temperature-dependent tidal $Q$ and a CTL one, the latter of which converges to a CPL model for small eccentricities as the time-lag of the principal tide becomes $1/Q$ \citep[][]{heller11}. For the CTL model we use the same framework and parameterization as in the work of \citet{heller13}, which goes back to \citet{leconte10} and \citet{hut81}. In particular, we use the following parameters: $k_2 = 0.3$ \citep[as in][]{henning09, heller13}, and tidal time lag, $\tau = 638 \mathrm{s}$, which was measured for the Earth \citep{lambeck77, neron97}.

In this model, the tidal heating function is linear in both $\tau \sim 1/(nQ)$ and $k_2$. Hence, changes in these parameters usually do not have dramatic effects on the tidal heating. In contrast, changes in the radius or mass of the tidally distorted object, or in the orbital eccentricity or semi-major axis result in significant changes of the tidal heating rates.

For the viscoelastic tidal heating calculations, we use the model described by \citet{dobos15}, which was originally developed by \citet{henning09}. Tidal heat flux is calculated from

\begin{equation}
   \label{viscel}
       h_\mathrm{s,visc} = - \frac {21} {2} Im(k_2) \frac {R_\mathrm{s}^5 n_\mathrm{s}^5 e_\mathrm{s}^2} {G} \, ,
\end{equation}

\noindent where $Im(k_2)$ is the complex Love number, which describes the structure and rheology in the satellite \citep{segatz88}. In the Maxwell model, the value of \textit{Im}($k_2$) is given by \citep{henning09}

\begin{equation}
   \label{Imk2}
       - Im(k_2) = \frac {57 \eta \omega} { 4 \rho g R_\mathrm{m} \left[ 1 + \left( 1 + \dfrac { 19 \mu } { 2 \rho g R_\mathrm{m} } \right)^2 \dfrac { \eta^2 \omega^2 } { \mu^2 } \right] } \, ,
\end{equation}

\noindent where $\eta$ is the viscosity, $\omega$ is the orbital frequency, and $\mu$ is the shear modulus of the satellite. The temperature dependency of the viscosity and the shear modulus is described by \citet{fischer90} and \citet{moore03}; and the adapted values and equations are listed by \citet{dobos15}. 
Since only rocky bodies like the Earth are considered as satellites in this work, the solidus and liquidus temperatures, at which the material of the rocky body starts melting and becomes completely liquid, were chosen to be 1600 K and 2000 K, respectively. We assume that disaggregation occurs at 50\% melt fraction, which implies a breakdown temperature of 1800 K.

The viscoelastic tidal heating model also describes the convective cooling of the body. The iterative method described by \citet{henning09} was used for our calculations of the convective heat loss:

\begin{equation}
   \label{qBL}
       q_\mathrm{BL} = k_\mathrm{therm} \frac {T_\mathrm{mantle} - T_\mathrm{surf}} {\delta(T)} \, ,
\end{equation}

\noindent where $k_\mathrm{therm}$ is the thermal conductivity ($\sim 2 \mathrm{W/mK}$), $T_\mathrm{mantle}$ and $T_\mathrm{surf}$ are the temperatures in the mantle and on the surface, respectively, and $\delta(T)$ is the thickness of the conductive layer. We use $\delta(T)=30 \, \mathrm{km}$ as a first approximation, and then for the iteration

\begin{equation}
   \label{delta}
       \delta(T) = \frac {d} {2 a_2} \left( \frac {Ra} {Ra_\mathrm{c}} \right)^{-1/4}
\end{equation}

\noindent is used, where $d$ is the mantle thickness ($\sim 3000$~km), $a_2$ is the flow geometry constant ($\sim 1$), $Ra_\mathrm{c}$ is the critical Rayleigh number ($\sim 1100$) and $Ra$ is the Rayleigh number which can be expressed as

\begin{equation}
   \label{Ra}
       Ra = \frac { \alpha \, g \, \rho \, d^4 \, q_\mathrm{BL} } { \eta(T) \, \kappa \, k_\mathrm{therm} } \, .
\end{equation}

\noindent
with $\alpha$ ($\sim 10^{-4}$) as the thermal expansivity, $\kappa = k_\mathrm{therm} / ( \rho \, C_\mathrm{p} )$ as the thermal diffusivity, and $C_\mathrm{p} = 1260 \, \mathrm{J/(kg \, K)}$. The iteration of the convective heat flux ends when the difference of the last two values becomes smaller than $10^{-10} \mathrm{W/m^2}$. Once the stable equilibrium temperature is found, we compute the tidal heat flux.

This viscoelastic model was already used together with a climate model by \citet{forgan16} with the aim of determining the location and width of the circumplanetary habitable zone for exomoons. The 1D latitudinal climate model included eclipses, the carbonate-silicate cycle and the ice-albedo feedback of the moon (in addition to tidal heating, stellar and planetary radiation). The ice-albedo positive loop in the climate model along with eclipses result in a relatively close-in outer limit for circumplanetary habitability, if the orbit of the satellite is not inclined. The climate model, however, can only be used for bodies of similar sizes to the Earth. In this work we investigate the effect of the viscoelastic model on smaller exomoons, as well, and for this reason, instead of a climate model, we apply an orbit-averaged illumination model. Beside solar-like host stars, we made calculations also for M dwarfs.

Both the CPL and the viscoelastic models are valid only for small orbital eccentricities,
that is, for $e{\lesssim}0.1$. For larger eccentricities, the instantaneous tidal heating in
the deformed body can differ strongly from the orbit-averaged tidal heating rate and
the frequency spectrum of the decomposed tidal potential could involve a wide range
of frequecies \citep{greenberg09}. Both aspects go beyond the approximations
involved in the models, and hence our results for $e{\gtrsim}0.1$ need to be taken with
a grain of salt.

\subsection{Simulation setup}

We consider various reference systems of a star, a planet, and a moon.

\subsubsection{Comparison of equilibrium and viscoelastic tides}
\label{sec:comparison}

First, we examine the different effects of tidal heating in either a CTL or in a viscoelastic tidal model on the location of the circumplanetary habitable edge (see Section~\ref{habedge}). We consider a star with a sun-like radius ($R_\star~=~R_\odot$) and effective temperature ($T_{\rm eff}~=~5778$\,K), a 5 Jupiter-mass gas giant at 1\,AU orbital distance with an eccentricity of 0.1, and a $0.5\,M_\oplus$ moon. The orbital period and eccentricity of the moon are varied between 1 and 20 days and between 0.01 and 1, respectively.

\subsubsection{Viscoelastic tides beyond the stellar habitable zone}

Second, we map the circumplanetary habitable zone over a wide range of circumstellar orbits (see Section~\ref{tidalHZ}) using four different star-planet-moon configurations. The star is either sun-like (see Section~\ref{sec:comparison}) or an M3 class main sequence star \citep[$M_* = 0.36 \, M_\odot$, $R_* = 0.39 \, R_\odot$, $T_\mathrm{eff} = 3250 \, \mathrm{K}$ and $L_* = 0.0152 \, L_\odot$,][Table 1]{kaltenegger09}. The planet is a Jupiter-mass gas giant, and the satellite's mass is either $0.1\,M_\oplus$ or $1\,M_\oplus$ (i.e. a Mars or Earth analogue). The density of the moon is that of the Earth. The stellar luminosity and the temperature values are used for our calculations of the HZ boundaries, and the stellar radius and temperature values are required for the incident stellar flux calculation.

\section{Results} \label{results}

\subsection{The habitable edge} \label{habedge}

\begin{figure}
	\centering   
	\includegraphics[width=21pc]{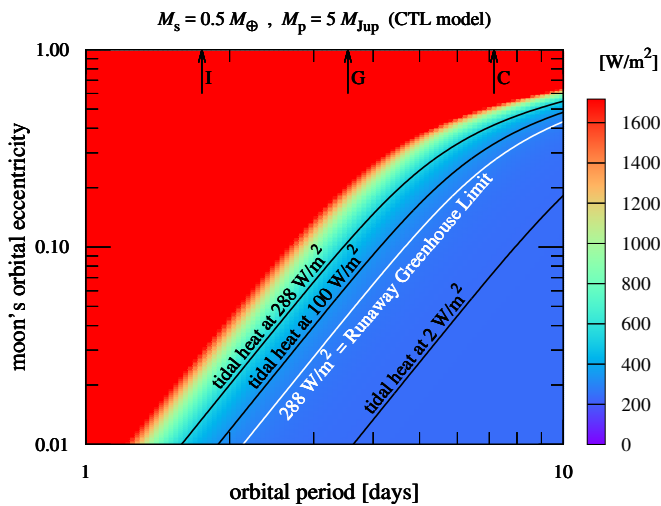}     
	\caption{\label{fixed}Energy flux at the top of the atmosphere of a 0.5 Earth-mass moon orbiting a 5 Jupiter-mass planet at 1 AU distance from a Sun-like star; tidal heating is calculated with a CTL model. The planetary orbit has 0.1 eccentricity. The stellar radiation, the planetary reflectance, thermal radiation of the planet and tidal heating were considered as energy sources. White contour curve indicates the runaway greenhouse limit considering all energy sources, while the black curves show the tidal heating flux alone. The arrows at the top of the figure indicate the orbital periods of Io (I), Ganymede (G) and Callisto (C). Note that the CTL model is valid only for small orbital eccentricities. Heating contours in the upper half of the panel (above 0.1 along the ordinate) could be off by orders of magnitude in real cases.}
\end{figure}

\begin{figure}
	\centering   
	\includegraphics[width=20pc]{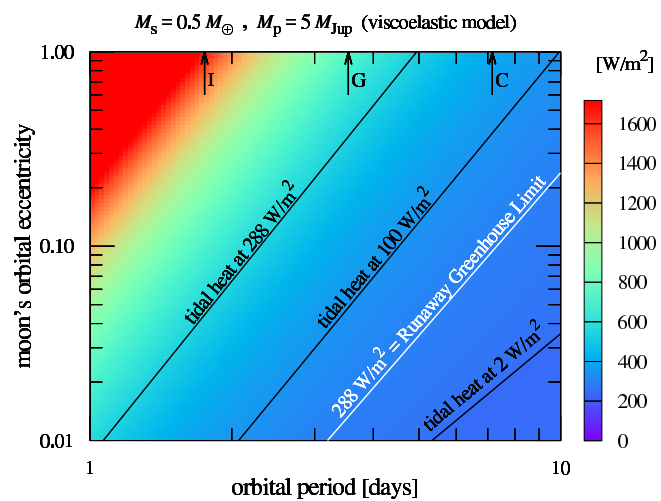}     
	\caption{\label{visc}Energy flux at the top of the atmosphere of a 0.5 Earth-mass moon orbiting a 5 Jupiter-mass planet at 1 AU distance from a Sun-like star; tidal heating is calculated with a viscoelastic model. The same colours, contours and signatures were applied as in Fig.~\ref{fixed}. Note that the viscoelastic model is valid only for small orbital eccentricities. Heating contours in the upper half of the panel (above 0.1 along the ordinate) could be off by orders of magnitude in real cases.}
\end{figure}

We calculated the total energy flux (using Eq.\,\ref{Fglob}) at the top of the moon's atmosphere as a function of distance to the planet in both the CTL (Fig.~\ref{fixed}) and the viscoelastic (Fig.~\ref{visc}) frameworks. In both Figs.~\ref{fixed} and \ref{visc}, colours show the amount of the total flux received by the moon, and the white contours at $288 \, \mathrm{W/m^2}$ indicate the habitable edge defined by the runaway-greenhouse limit \citep{heller13}. Interestingly, in the CTL model the habitable edge is located closer in to the planet than in the viscoelastic model.

Black contour curves show the tidal heat flux alone at $288$, $100$ and $2 \, \mathrm{W/m^2}$, the latter is the global mean heat flow from tides on Io as measured with the Galileo spacecraft \citep{spencer00}. This curve in Fig.~\ref{visc} has a different slope than the one labeled `tidal heat at $100 \, \mathrm{W/m^2}$'. The different slope is caused by the changing equations in the viscoelastic model. The tidal flux (and also the convective cooling flux) is described by different formulae below or above certain temperatures (namely the solidus, the breakdown and the liquidus temperatures). In other words, if the tidal heating is low, different equations will be used, than in the case of higher tidal forces.

Figs.~\ref{fixed} and \ref{visc} show substantial differences. The viscoelastic model predicts moderate ($\leq~2 \, \mathrm{W/m^2}$) heating beyond $5.3$ days of an orbital period for $e = 0.01$, while the CTL model needs the moon to be as close as $3.5$ days to generate the same tidal heating with the same orbital eccentricity. In other words, the viscoelastic model predicts significant tidal heating in wider orbits. At close orbits, however, the CTL model products extremely high fluxes. With a 1.6 day orbital period (similar to Io, see arrows at the top), our test moon with e = 0.01 would generate a tidal heat flux that is sufficient to trigger a runaway greenhouse effect. In this case, stellar illumination is not even required to make such a moon uninhabitable. In contrast, with the viscoelastic model the moon would need to have a 1 day orbital period to trigger the same effect.

A comparison of these two plots shows the `thermostat effect' of the viscoelastic model described by \citet{dobos15}. The CTL model yields lower tidal heating rates than the viscoelastic model in the weak-to-moderate heating regime (tidal heat $\leq 100 \, \mathrm{W/m^2}$), while above $100 \, \mathrm{W/m^2}$ the viscoelastic model produces lower heating rates, where the CTL model runs away. As a consequence, in this specific example of a 0.5 Earth-mass moon around a 5 Jupiter-mass planet, the habitable edge (red contour curve in the figures) is located at a larger distance from the planet than for the CTL model. It is caused by the stronger tidal heating rate below $100 \, \mathrm{W/m^2}$.

Stellar and planetary illuminations can also have important effects. They can break the feedback loop between tidal heating and convective cooling: in an extreme case the system will not even find a stable convective heat transport rate. However, the most relevant cases of tidally heated exomoons will involve systems orbiting so far from the star that illumination heating is irrelevant and around planets so old that planetary illumination is also small. In any cases, such systems will be the ones that can be most easily understood, if they exist.

\subsection{Circumplanetary tidal habitable zone} \label{tidalHZ}

\begin{figure*}
	\centering   
	\includegraphics[width=40pc]{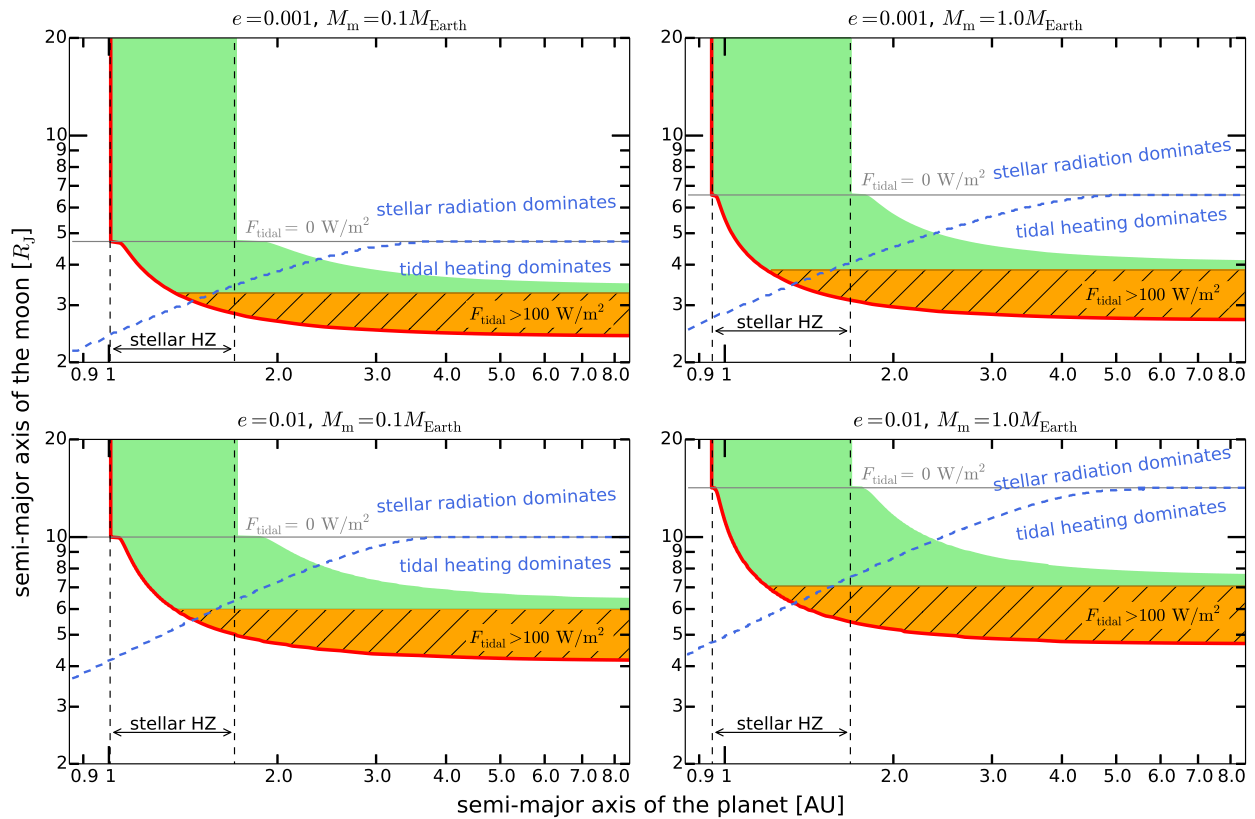}     
	\caption{\label{G2}Tidal heating habitable zone for 0.1 and 1 Earth-mass exomoons (left and right panels, respectively) around a Jupiter-mass planet hosted by a G2 star as functions of the semi-major axes of the planet and the moon. The red curve indicates the runaway greenhouse limit (habitable edge). The green and the diagonally striped orange areas cover the habitable region, where the striped orange colour indicates that the tidal heating flux is larger than $100 \, \mathrm{W/m^2}$. The orbital eccentricity of the moon is 0.001 in the top panels, and 0.01 in the bottom panels. Vertical dashed lines indicate the inner and outer boundaries of the circumstellar habitable zone. At the dashed blue contour curve the tidal heating flux equals to the stellar radiation flux, hence it separates the tidal heating dominated and the stellar radiation dominated regime. At the grey horizontal line the tidal flux is zero.}
\end{figure*}

\begin{figure*}
	\centering   
	\includegraphics[width=40pc]{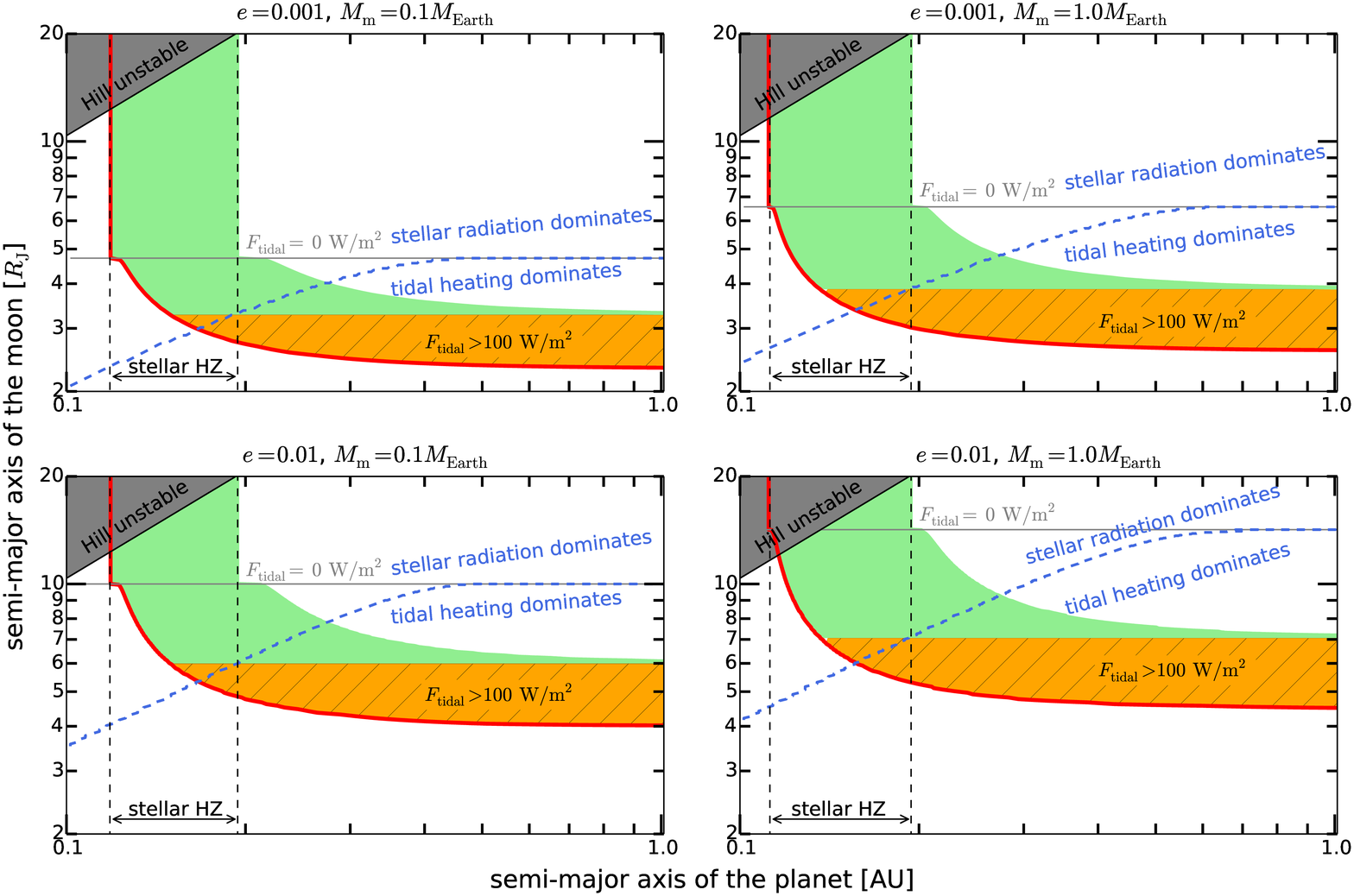}     
	\caption{\label{M3}Tidal heating habitable zone for 0.1 and 1 Earth-mass exomoons (left and right panels, respectively) around a Jupiter-mass planet hosted by an M3 dwarf as functions of the semi-major axes of the planet and the moon. The orbital eccentricity of the moon is 0.001 in the top panels, and 0.01 in the bottom panels. The same colours and contours were applied as in Fig.~\ref{G2}. The grey area in the upper left corner covers those cases where the moon's orbit is considered unstable.}
\end{figure*}

In addition to the inner habitable edge, we also want to locate the outer boundary of the circumplanetary habitable zone for different satellite sizes and stellar classes. Figs.~\ref{G2} and \ref{M3} show the circumplanetary habitable zones for our Jupiter-like test planet with a moon around either a sun-like or an M dwarf star, respectively. The left and right panels of the figures show the cases of 0.1 and 1 Earth-mass satellites, respectively. The eccentricity of the moon's orbit is 0.001 in the top panels and 0.01 in the bottom panels.

Grey horizontal lines in each panel indicate zero tidal heat flux. Above these lines, only stellar radiation is relevant since the reflected and thermal radiation from the planet are also negligibly small. Green areas illustrate habitable surface conditions with tidal heating rates below $100\,{\rm W\,m}^{-2}$, whereas diagonally striped orange areas visualize exomoon habitability with tidal heating rates above $100\,{\rm W\,m}^{-2}$ \citep[color coding adopted from][]{2014AsBio..14...50H}.

As expected, for smaller satellite masses the HZ is located closer to the planet. At zero tidal flux, small plateaus are present which are consequences of the viscoelastic tidal heating model. At the horizontal line tidal forces are turned on, hence there is a change in the position of the green area. At the plateau the tidal heat flux elevates from zero to about $10 \, \mathrm{W/m^2}$ or more. The sudden change is caused by the fact that the tidal heating model gives result only if the tidal heat flux and the convective cooling flux has stable equilibrium. It means that if tidal forces are weak, then the convective cooling will be weak too, or not present in the body at all, hence there is no equilibrium. In other words, tidal heating is insufficient to drive convection in the body. There will still be some heating, but a different heat transport mechanism (probably conduction) would be in play. A more accurate model would consider all heat transport mechanisms at all temperatures and would probably not exhibit such discontinuities.

In the inspected cases the two dominating energy sources are the stellar radiation and the tidal heating. Since these two effects are physically independent, one would not expect them to be of comparable importance except in a small fraction of cases. In general, however, one effect will dominate the other. To explore this interplay between stellar illumination and tidal heating in more detail, we calculated the distance from the star at which tidal heating (for a given hypothetical moon and orbit) equals stellar heating. These critical distances are indicated with blue dashed curves in  Figs.~\ref{G2} and \ref{M3}. We find that beyond 3-5\,AU around a G2 star and beyond 0.4-0.6\,AU around an M3 star, tidal heating for moons with eccentricities between 0.001 and 0.01 will be the dominant source of energy rather than stellar radiation (see the overlapping of the grey and dashed blue curves). This is the case for both satellite masses investigated. For smaller stellar distances, the dominant heat source depends on the circumplanetary distance of the moon.

Based on the fact that Io is a global volcano world, one might take the $2 \, \mathrm{W/m^2}$ limit as a conservative limit for Earth-like surface habitability. But other worlds might still be habitable even in a state of extreme tidal heating near $100 \, \mathrm{W/m^2}$. In fact, the surface habitability of rocky, water-rich planets or moons has not been studied in this regime of extreme tidal heating to our knowledge. Nevertheless, although $100 \, \mathrm{W/m^2}$ seems a lot of internal heating compared to the internal heat flux on Earth, which is  less than $0.1 \, \mathrm{W/m^2}$ \citep{2007SSRv..129...35Z}, it could still allow the presence of liquid surface water as long as the total energy flux is below the runaway greenhouse limit.  The striped orange colour indicates that the body might become volcanic or tectonically active in this regime.

It is supposed that the gravitational pull of an M dwarf can force an exomoon whose orbit is within the stellar HZ into a very eccentric circumplanetary orbit \citep{heller12}. The resulting tidal heating might ultimately prevent such moons from being habitable. Since the planetary Hill sphere in the stellar HZ of M dwarfs is not so large, any moon needs to be in a tight orbit. Prograde (regular) moons are only stable out to about 0.5 Hill radius \citep{domingos06}. In Fig.~\ref{M3}, the area beyond this line is shown as a shaded region in grey and labeled as `Hill unstable'. As a consequence of the closeness of this Hill unstable limit, the circumplanetary volume for habitable satellites is much smaller in systems of M dwarfs than in systems with sun-like stars, in particular if the moons have substantial eccentricities. Note how the circumplanetary space between the red (runaway greenhouse) curve and the Hill unstable region becomes smaller as the moon's eccentricity or its mass increases.

\section{Discussion}

From the results shown in Section \ref{results}, the following general findings can be concluded.

The circumplanetary habitable edge calculated with the viscoelastic model (Fig.~\ref{visc}) is located at a larger distance than in the CTL model (Fig.~\ref{fixed}). If the outer boundary of the circumplanetary habitable zone is defined by Hill stability, than it means that the viscoelastic model reduces the habitable environment. However, \citet{forgan14} showed that the moon can enter into a snowball phase, if the ice-albedo feedback and eclipses are also taken into account. It means that the outer boundary will be significantly closer to the planet, than in the case when it is defined by Hill stability. Since the viscoelastic model resulted in higher tidal fluxes below $100 \, \mathrm{W/m^2}$, than the CTL model, we expect that the outer boundary defined by the snowball state will be farther from the planet. Altogether, the circumplanetary HZ is not likely to be thinner in the viscoelastic case, but it is located in a larger distance from the planet.

Tidal heating in close-in orbits is more moderate with the viscoelastic model and does not generate extremely high rates as predicted by the CTL model.

Moons that are primarily heated by tides, that is, moons beyond the stellar habitable zone, have a wider circumplanetary range of orbits for habitability with the viscoelastic model than with the CTL model. This is in agreement with the findings of \citet{forgan16}. In \citet[Fig.~2]{2014AsBio..14...50H}, it was shown that the circumplanetary habitable zone calculated with the CTL model thins out dramatically as the planet-moon binary is virtually shifted away from their common host star. This is due to the very strong dependence of tidal heating in the CTL model on the moon's semi-major axis.

In the calculations we considered 0.1, 0.5 and 1 Earth-mass moons. \citet{lammer14} found that about 0.25 Earth masses (or 2.5 Mars masses) are required for a moon to hold an atmosphere during the first 100 Myrs of high stellar XUV activity, assuming a moon in the habitable zone around a Sun-like star. Considering this constraint, a 0.1 Earth-mass moon may not be habitable in the stellar HZ because of atmospheric loss, however, at larger stellar distances the stellar wind and strong stellar activity do not present such severe danger to habitability.

From a formation point of view, smaller moons are more probable to form around super-Jovian planets than Earth-mass satellites, according to \citet{canup06}, who gave an upper limit of $10^{-4}$ to the mass ratio of the satellite system and the planet. However, moons can also originate from collisions rather than the circumplanetary disk, as in the case of the Moon. In such cases, larger mass ratios are reasonable, too.

We have considered a large range of orbital eccentricities for the moon with values up to 1. Beyond the fact that our models are physically plausible only for moderate values ($\leq 0.1$), tidal circularization will act to decrease eccentricities to zero on time scales that are typically much shorter than 1 Gyr  \citep{2011ApJ...736L..14P, heller13}. Hence, even moderate eccentricities can only be expected in real exomoon systems if the star has a significant effect on the moon's orbit \citep{heller12}, if other planets can act as orbital perturbers \citep{2013ApJ...769L..14G, 2013ApJ...775L..44P, 2015MNRAS.449..828H}, if other massive moons are present around the same planet, or if the planet-moon system has migrated through orbital resonances with the circumstellar orbit \citep{2010ApJ...719L.145N, 2016ApJ...817...18S}.

Hot spots (hot surface areas generated by geothermical heat, that could generate volcanoes) might be important on both observational and astrobiological grounds. Regarding observations, hot spots produce variability in the brightness and spectral energy distribution of thermal emission from the moon. As Io illustrates, hot spots can potentially be big and prominently hot \citep[see][Fig.~4]{spencer00}. This could let one determine the moon's spin period and may indicate a geologically `active' body. Hot spots also allow liquid water to locally exist and persist over long periods of time even if the mean surface temperature is far below the freezing point. Enceladus may provide an example for such phenomenon in the Solar System, where tidal heating maintains a liquid (and probably global) ocean below the ice cover, and contributes to the eruption of plumes at the southern region of the satellite \citep{thomas16}.

On the other hand, hot spots can be effective in conducting the internal heat. The calculated fluxes in this paper are average surface fluxes, but in reality hot spots are probable to form. These areas are much higher in temperature, meaning that other areas on the surface must be somewhat colder than the average. As a result, the temperature of the surface (excluding the vicinity of the hot spots) can be lower than that is calculated.

Two end-member models exist for spatial distribution of tidal heat dissipation. In model A the dissipation mostly occurs in the deep-mantle of the body, and in model B it occurs in the asthenosphere, which is a thin layer in the upper mantle \citep{segatz88}. In model A the surface heat flux is higher at the polar regions, while in model B the flux is mostly distributed between $-45\,^{\circ}$ and $+45\,^{\circ}$ latitudes. From volcano measurements of Io it seems that model B is more consistent with the location of hot spots \citep{hamilton11, hamilton13, rathbun15}. The global distribution of volcanoes on the surface is random (Poisson distribution), but closer to the equator they are more widely spaced \citep[uniform distribution,][]{hamilton13}.

On Io the total power output of volcanoes and paterae is about $5 \cdot 10^{13} \, \mathrm{W}$ \citep{veeder12, veeder12b}. About 62\% of the flux is going through volcanoes in Io \citep{veeder12}, which means that there is a large temperature difference in the volcanic areas and lowlands. For moons of larger sizes than Io, or with much stronger tidal force, it is probable that even higher percentage of energy will leave through volcanoes. A geological model is needed to estimate the connection between the tidal heating flux and the energy output of volcanoes and hot spots, which is beyond the scope of this paper.

Detectibility of Mars-to-Earth size satellites is not investigated in this work, but detections could be possible in the \textit{Kepler} data \citep{kippingetal09,heller14a} or with the PLATO \citep{2015ApJ...810...29H} or CHEOPS \citep{2015PASP..127.1084S} missions, or with the E-ELT \citep{2014ApJ...796L...1H,2016A&A...588A..34H,2016ApJ...824...76S}. Naturally, larger moons are easier to detect than smaller moons, and so we expect an observational selection bias for the first known exomoons to be large and potentially habitable.

\section{Conclusion}

In this work we improved the viscoelastic tidal heating model for exomoons \citep{dobos15} by adding the stellar radiation, the planetary reflectance and the planet's thermal radiation to the energy budget. We found that the `thermostat effect' of the viscoelastic model is robust even with the inclusion of these additional energy sources. This temperature regulation in the viscoelastic tidal heating model is caused by the melting of the satellite's inner material, since the phase transition prevents the temperature from rising to extreme heights.

We investigated the circumplanetary tidal heating HZ for a few representative configurations. We showed that the extent of the tidal HZ is considerably wide even at large distances from the stellar HZ, predicting more habitable satellite orbits than the CTL models. In a previous study with a CTL model, \citet{2014AsBio..14...50H} found that the tidal HZ thinned out for large stellar distances. We also showed that if tidal heating is present in the moon, then beyond 3--5~AU distance from solar-like stars this will be the dominating energy source, and for M3 main sequence dwarfs tidal heating dominates beyond 0.4--0.6~AU distance. At smaller stellar distances the semi-major axes of the moon defines whether stellar radiation or tidal heating dominates.

From an observational point of view, M dwarfs are better candidates to detect habitable exomoons, since both the stellar and the circumplanetary tidal HZs and closer to the star, meaning that the orbital period of the planet is shorter, hence more transits can be observed. However, in the stellar HZ, the possible habitable orbits for exomoons are constrained by Hill stability. At larger stellar distances the circumplanetary tidal HZ and the Hill unstable region do not overlap, so there is no such constraint.

\section*{Acknowledgements}
We thank Duncan Forgan for a very helpful referee report. VD thanks L\'aszl\'o L. Kiss for the helpful discussion. This work was supported in part by the German space agency (Deutsches Zentrum f\"ur Luft- und Raumfahrt) under PLATO grant 50OO1501. This research has been supported in part by the World Premier International Research Center Initiative, MEXT, Japan. VD has been supported by the Hungarian OTKA Grants K104607, K119993, and the Hungarian National Research, Development and Innovation Office (NKFIH) grant K-115709.

\bibliographystyle{apalike}
\bibliography{ref}

\end{document}